\documentclass[12pt]{iopart}
\usepackage{amssymb,graphicx}
\begin{document}

\title[Time dependent variational principle]{Time Dependent Variational Principle and Coherent State Orbits for a Trapped Ion} 
\author{Bogdan M. Mihalcea}

\address{Institute for Laser, Plasma and Radiation Physics (INFLPR), 
\newline Atomi\c stilor Str. Nr. 409, 
\newline 077125 M\u agurele-Bucharest, Jud. Ilfov, Romania}
\eads{\mailto{bogdan.mihalcea@infim.ro}}
\begin{abstract}

Spectral properties of the Hamiltonian function which characterizes a trapped ion are investigated. In order to study 
semiclassical dynamics of trapped ions, coherent state orbits are introduced as sub-manifolds of the quantum state space, with 
the K\"ahler structure induced by the transition probability. The time dependent variational principle is applied on coherent 
states' orbits. The Hamilton equations of motion on K\"ahler manifolds of the type of classical phase spaces naturally arise. 
The associated classical Hamiltonian is obtained from the expected values on symplectic coherent states of the quantum 
Hamiltonian. Spectral information is thus coded within the phase portrait. We deal with the bosonic realization of the Lie 
algebra of the SU(1,1) group, which we particularize for the case of an ion confined in a combined, Paul and Penning trap. This 
formalism can be applied to Hamiltonians which are nonlinear in the infinitesimal generators of a dynamical symmetry group, 
such as the case of ions confined in electrodynamic traps. Discrete quasienergy spectra are obtained and the corresponding 
quasienergy states are explicitly realized as coherent states parameterized by the stable solutions of the corresponding 
classical equations of motion. A correspondence between quantum and classical stability domains is thus established, using the 
Husimi representation.

\end{abstract}

\vspace{2pc}
\maketitle

\section{Introduction. Classical-quantum correspondence}
\label{intro}

The development of quantum mechanics and of its concepts left it with a heavy legacy of classical concepts. The most notable among them is Bohr's correspondence principle, which 
states that in the limit of large quantum numbers there are similarities between classical and quantum dynamics. In other words, the results from classical mechanics are 
macroscopically correct and may be regarded as limit cases of the quantum mechanics results, when quantum physics discontinuities are negligible. This principle is still 
used as an intuitive guide for finding quantum properties which are similar to known classical laws. Some of these analogies are intriguing, considering the fundamental differences 
in the mathematical formalism underlying the two theories: quantum mechanics uses a separable Hilbert space with a unitary inner product, while classical mechanics is based on a 
continuous phase space endowed with a symplectic structure \cite{Golds80, Peres02}. Nevertheless, in certain situations it might represent a weak correspondence between classical 
and quantum concepts. The concept of observables in quantum mechanics and in classical probability theory has drawn a lot of interest. It was shown that, by means of injective 
statistical maps, quantum mechanics can to a certain extent be reformulated in classical terms \cite{Stulpe1997}.

Pure and mixed states of a $n$-dimensional quantum entity can be represented as points of a subset of a $n^2$ dimensional real space \cite{Coecke1995}. Mathematically, a pure 
quantum state is typically represented by a vector in a Hilbert space. Mixed quantum states are characterized by means of density matrices. Both mixed and pure state are intensively 
investigated, due to a very large interest they present for modern physics. The expressions {\it coherent} and {\it incoherent} superposition of quantum states are also used in 
order to help distinguish between pure and mixed states. Superpositions of quantum states exhibit many features similar to those of their classical counterparts: they are the 
so-called coherent states, investigated by Schr\"odinger in 1926 \cite{Schro1926} and then rediscovered by Klauder \cite{Klaud1960, Klaud1963}, Glauber \cite{Glaub1963a, Glaub1963b} 
and Sudarshan \cite{Sudar1963} at the beginning of the 1960s. Glauber coined the term 'coherent states' in 1963, in the context of quantum optics \cite{Glaub1963c}. Such states are 
superpositions of Fock states of the quantized electromagnetic field which, below the limit of a complex factor, are not modified by the action of the photon annihilation operator 
\cite{Gazeau2009, Sanders2012}. Since then coherent states had a huge impact on quantum physics \cite{Perel72, Perel86, Feng1980, Bouwmeester2000, Dodon02, Leibf03, Grynberg2010} with 
widespread applications such as nuclear, atomic and condensed matter physics \cite{Horvath97, Sacke00, Vogel01, Fox2006, Haroche2006, Chen2007}, quantum field theory \cite{Orszag2008}, 
problems of quantization and dequantization \cite{Haroche2006, Chen2007, Combescure2012, Major2005}, non-commutative quantum mechanics \cite{BenGeloun2009} trying to unify quantum mechanics 
and gravity, path integrals, signal analysis and recently, quantum information processing (QIP) \cite{Fox2006, Haroche2006, Chen2007, Werth2009, Leibf97, Marzoli2009} using the 
entanglement feature \cite{Gazeau2009, Fox2006, Orszag2008, Leibf2005, Reichle2006}. Moreover, there are a lot of quantum states of interest for modern physics \cite{Sanders2012, Dodon2003} 
such as Fock states, Schr\"odinger-cat states \cite{Leibf03, Leibf2005} or squeezed states \cite{Holle1979, Caves80, Walls1983, Yuen1976}. A matter of interest is 
the engineering of quantum states showing negative parts in their Wigner functions, which is a signature for non-classical properties of the state that can not be explained using a 
classical phase-space distribution. Solutions of the linearized quantum Liouville equation for a plasma in a uniform magnetic 
field were obtained in the coherent-state basis and the linear response of such a system to general electromagnetic 
perturbations was investigated \cite{Nag1980}. Other plasma physics applications are identified in \cite{Groza2012a, Groza2012b}. 
All these problems are of large interest, as they discuss the relation between classical and quantum physics, of the 
still blurred separation line between them.          

Time-dependent variational method represents one of the most suited methods to characterize the time-evolution of quantum 
mechanical systems in the framework of a possible approximation. The quantum enginnering mechanism on which the method is based 
consists in preparing a trial state for the variation as a function of variational parameters.  
If a certain condition is introduced, the time-dependent variational method can be formulated in the framework of classical 
Hamiltonian mechanics \cite{Ceausescu1987}. Moreover, the time-dependent Hartree-Fock (TDHF) is one of the time-dependent 
mean-field frameworks in the nuclear many-body theories and has been made use of in describing the microscopic 
mechanism of nuclear collective motions which are represented as classical trajectories in the TDHF phase space. 
Since the TDHF phase space is the classical correspondent of the quantum space of states ( i.e., a coherent state representation 
of the boson-mapped fermion space of states), classical informations obtained from the TDHF trajectories supplies us with 
relevant data on the structure change which occurs in the quantum space of states \cite{Hashimoto2001}. 
Time-dependent Hartree-Fock (TDHF) approximation of single-particle dynamics in systems of interacting fermions was studied in \cite{Bardos2004}. The time-dependent Hartree–Fock 
(TDHF) method was used to simulate the behavior of the electronic density prior to ionization, for molecules in high intensity oscillating electric fields \cite{Li2004}. A canonical Hamiltonian 
formulation for the general time-dependent variational principle associated with the Schr\"odinger equation is discussed in \cite{Kerman1976}. Such methods have proven to be useful 
when studying a wide range of systems including the many body problem and field theory.

A charged particle driven by a time-dependent perturbation in a quantum system is a non-trivial fundamental issue, with applications such as nonlinear dynamics \cite{Porter2001}, 
quantum simulation of the Dirac equation and implicitly quantum field theories \cite{Lamata2007, Gerritsma2010}, cavity QED \cite{Barros2009, Russo2009}, quantum non-locality 
\cite{Blatt2008}, quantum transport \cite{Tang99}, quantum optics \cite{Grynberg2010, Faria96, Vogel93, Vogel2006}, superpositions of arbitrary quantum states \cite{Kish03} or 
quantum information processing \cite{Sanders2012, Leibf2005, Reichle2006, Haffner2005, Benhelm2008, Monz2009, Home2011}. The paper investigates quantum dynamics for systems with dynamical 
symmetry, such as is the case of ions confined in electrodynamic traps or combined (Paul and Penning) traps \cite{Ghosh95}. The coherent states' orbits are introduced as 
sub-manifolds of the quantum state space, with a K\"ahler structure induced by the transition probability. An algorithm is proposed, which associates a classical Hamiltonian to the 
quantum Hamiltonian describing the ion. The classical Hamiltonian implicitly contains the spectral information of the quantum system. The coherent states formalism introduced in 
\cite{Gheor92, Gheor2000} and developed in \cite{Major2005} is used, and the time dependent variational principle (TDVP) \cite{Feng1980, Krame81, Bozic91, Drago04}, in order to 
characterize the semiclassical dynamics.

The results have been applied to the case of an ion confined in a quadrupole electromagnetic trap with anharmonicities. The coherent states formalism for dynamical groups and the 
time dependent variational principle (TDVP) are applied with an aim to investigate semiclassical behaviour of trapped ions. The explicit expression of the quantum Hamiltonian 
associated to a charged particle is found, as a function of the Lie algebra generators for the radial and axial symplectic groups. The Schr\"odinger equation solution is also 
obtained. The Hamiltonian function is then particularized to the case of combined, quadrupole and octupole traps.  

The paper is organized as follows: Section \ref{CSDSG} reviews some definitions from the geometry of coherent states. In Section \ref{TDVP} the time dependent variational 
principle is applied to a manifold of test vectors in paragraph \ref{VPTS}. The Lagrange function is introduced and the Hamilton equations of motion result for the system we have 
investigated in Section \ref{Lagrange}. Section \ref{Symplectic} presents the coherent states built over weight states of the discrete positive series of the symplectic group 
${\mathcal Sp}\left( 2,{\mathbb R}\right) $ and the Lie algebra generators. A model is proposed which enables explicit calculus of the Husimi function for every algebraic model 
with a dynamical group. The equations of motion in phase space arise. The model we suggest is applied in Section \ref{Quadyn} to a combined Paul and Penning trap, with octupole 
anharmonicity. The Lie algebra generators for the radial and axial symplectic group ${\mathcal Sp}\left( 2,{\mathbb R}\right)_{a,r} $ are introduced in Section 
\ref{AlgebraicModels}, together with the Hamiltonian of the system. The symplectic coherent states representation is introduced in Section \ref{SymplecRepr}. The 
classical Hamiltonian for the system we investigated finally results. Section \ref{Concl} is dedicated to a discussion of the results.           

\section{Coherent states for dynamical symmetry groups}
\label{CSDSG}

\subsection{Notions of differential geometry}
Euclidian space is a basic symplectic manifold not only because it is simple to characterize, but mainly owing to the fact that it provides a local model for every symplectic 
manifold. The Euclidean space can be treated as a semi-local symplectic manifold. Hence we can submit global nonlinear problems on it. The phase space exhibits a geometric 
structure, as it is a space endowed with a tensor. Such space is called a symplectic form. Symplectic manifolds are even-dimensional. They are studied using symplectic geometry or 
symplectic topology \cite{Fomen95, McDuff98}, which represents a branch of differential geometry (topology) \cite{Jost2009}. Symplectic manifolds arise naturally in abstract 
formulations of classical mechanics and analytical mechanics as the cotangent bundles of manifolds, e.g. in the Hamiltonian formulation of classical mechanics \cite{Fomen95, 
Bolsi2004}. The subject of symplectic topology is the global structure of a symplectic manifold and the behaviour of symplectomorphisms which are far from identity. These aspects 
are of interest both in Euclidean space with its associated standard linear symplectic structure and on general symplectic manifolds \cite{McDuff98, Jost2009}. A K\"ahler manifold 
is a symplectic manifold with an integrable almost complex structure. A consequence of the integrability theorem is the fact that every Riemann surface carries a K\"ahler structure 
\cite{Salvai2005}. Most symplectic manifolds are not K\"ahler, which means they do not possess an integrable complex structure compatible with the symplectic form. Gromov 
\cite{Gromo82, Gromo99} made the important observation that symplectic manifolds admit an abundance of compatible almost complex structures, therefore they satisfy all the axioms 
for a K\"ahler manifold except the requirement that the transition functions be holomorphic. Differential geometry has applications to both Lagrangian mechanics and Hamiltonian 
mechanics. Symplectic manifolds in particular can be used to study Hamiltonian systems. Any real valued differential function $H$ on a symplectic manifold can serve as an 
energy function or Hamiltonian.

\subsection{Coherent states geometry} 
Further on, we will recall some definitions from the geometry of coherent states \cite{Perel72, Perel86, Werth2009}, while introducing the elementary quantum systems (in the Wigner 
sense), which represent the most extended class of models describing both analitically and numerically the energetic spectra and the solutions of the Schr\"odinger equation. 

Let $G$ be a Lie group acting on a manifold $M$ (respectively, a complex Lie group acting on a complex manifold $M$). Then for every point $m \in M$ we define its {\it orbit} by 
${\mathcal O}_m = G_m = \{g.m\arrowvert g\in G\}$ and stabilizer by 
\cite{Kirillov2008}
\begin{equation}
G_m=\{ g \in G\arrowvert g.m=m\} 
\end{equation}  

We will denote by $\mathcal H$ the separable complex Hilbert space of the state vectors (quantum states). The state space in quantum mechanics will be denoted by 
${\mathcal P}({\mathcal H})$. If $\psi \in {\mathcal H}$ is a state vector, then $\widehat{\psi }=\left\{ \lambda \psi |\lambda \in {\mathbb C}\right\} $ represents the state 
associated to $\psi $. Then 
\begin{equation}
{\mathcal P}(\mathcal H)=\left\{ \hat \psi \ \Big|\ \psi \in \mathcal H\backslash \left\{ 0\right\} \right\}\;,
\end{equation}
is the state space, which is also a projective complex space. $\widehat{\psi }$ is an unidimensional, complex linear space. If using normalized states 
$\left(\left\| \psi \right\| =1\right)$, we can introduce the states as radii and we have $\widehat{\psi }=\left\{ e^{i\varphi }\cdot \psi \ \Big|\ \varphi \in {\mathbb R}\right\} $. 
Further below we will consider ${\mathcal U}$ as an unitary irreducible representation (UIR) of the Lie group $\mathcal G$ in $\mathcal H$. This means that $\mathcal U(g)$ is a unitary 
operator on ${\mathcal H}$ for $\forall g \in {\mathcal G}$, such as
$$
{\mathcal U}\left(g \cdot g'\right) = {\mathcal U}\left(g\right) \cdot {\mathcal U}\left(g'\right),\ \forall g, g' \in {\mathcal G}
$$
$$
{\mathcal U}(g^{-1})=\left[{\mathcal U}\left(g\right)\right]^{-1}={\mathcal U}^+\left(g\right) \;.
$$
where ${\mathcal U}^+$ is the adjoint of ${\mathcal U}$. This representation induces an action of the  group ${\mathcal G}$ on ${\mathcal P}({\mathcal H})$, defined as 
$g\hat \psi =\widehat{{\mathcal U}\left( g\right)\psi }$  for $\forall g \in {\mathcal G}$ and $\psi \in {\mathcal H}\backslash \{ 0 \}$. Then ${\mathcal G}_0$ is the {\it 
stationary group} of the state $\hat \psi $, defined as
$$
{\mathcal G}_0=\left\{g \Big|g\cdot \hat \psi =\hat \psi \ ,\;\ g\in {\mathcal G}\right\} 
$$
The orbit ${\mathcal G}\cdot \hat \psi =\left\{g\cdot \hat \psi \ \Big|\ g\in {\mathcal G}\right\} $ is called a coherent state orbit, while any state $g\cdot \hat \psi $ is called 
a {\it coherent state} for the group ${\mathcal G}$. We will denote $\mathcal O_{\hat \psi}= \mathcal G\cdot \hat \psi$. The {\it coherent vectors} which determine the coherent 
states, can be parametrized with the coordinates of the quotient space ${\mathcal M} = {\mathcal G}/{\mathcal G}_0$. For a large class of symmetry groups ${\mathcal G}$, 
${\mathcal M}$ presents a structure of classical phase space (symplectic manifold). Hence it is convenient to parametrize the try functions with points from ${\mathcal M}$. There 
exists a bijection between the points from ${\mathcal M}$, considered as classical states, and the coherent states. 

The triplet $\left( {\mathcal H}, {\mathcal U}, {\mathcal G}\right) $ is defined as an {\it {elementary quantum system}} in the Wigner sense or a {\it {quantum system with symmetry}}, 
where ${\mathcal H}$ is a separable Hilbert space, ${\mathcal G}$ is a group and ${\mathcal U}$ stands for an UIR. Coherent states can be associated with such systems. We will 
consider that the orbit ${\mathcal O}_{\hat \psi }$ is a sub-manifold of the space ${\mathcal P} \left({\mathcal H}\right)$ with a structure of K\"ahler space induced by the 
transition probability. Coherent state K\"ahler orbits are obtined for Heisenberg groups, compact groups and semisimple groups, with representations belonging to the holomorphic 
discrete series. The coherent states considered here have important applications in quantum physics and quantum optics, especially when referring to the physics of trapped ions.  

Thus a family of coherent states provides the opportunity of investigating the quantum world using the methods of study applied to the classical world. This feature allowed Glauber 
and others to treat a quantized boson or fermion field as a classical field, with an aim to compute correlation functions and other quantities relevant for statistical physics, such 
as partition functions. Therefore the system can be characterized by means of the associated classical trajectories.      

\subsection{SU(1,1) group in quantum optics}

We now turn our attention to elements of the group theory, especially those related to the $SU(1,1)$ group, the most elementary noncompact non-Abelian simple Lie group. It has 
several series of unitary irreducible representations (UIR): discrete, continuous and supplementary \cite{Bargmann1947, Vilenkin1968}. This paper refers only to the case of the 
discrete series, that exhibits many interesting physical applications \cite{Perel86}. The Lie algebra corresponding to the group $SU(1,1)$ is spanned by three operators 
${K_0,K_1,K_2}$, which satisfy the following commutation relationships \cite{Brif1997}
\begin{equation}\label{dinq1}
 [K_1, K_2] = -i K_0\, \ \ , [K_2, K_0] = i K_1\, \ \ , [K_0, K_1] = i K_2\ .
\end{equation}
We introduce the raising and lowering operators $K_{\pm} = K_1 \pm i K_2$ which satisfy
\begin{equation}\label{dinq2}
 [K_0, K_{\pm}] = \pm K_{\pm} \, \ \ [K_-, K_+] = 2 K_0
\end{equation}
The Casimir operator $K^2 = {K0}^2 - {K_1}^2 - {K_2}^2$ for any UIR can be expressed as the product between the identity operator $I$ and a number $k$: $K = k(k-1) I$. Therefore, 
representations of $SU(1,1)$ are determined by a single number $k$. In case of the discrete series representation, $k$ can have discrete values such as $k = \frac{1}{2}, 1, 
\frac{3}{2}, 2, \ldots$. The representation Hilbert space is spanned by the orthonormal basis $|k, n \rangle$, where $n = 0, 1, 2, \ldots \ $. A. O. Barut and L. Girardello 
\cite{Barut71}, then A. M. Perelomov \cite{Perel72, Perel86} have brought in the $SU(1,1)$ group to quantum optics. They have constructed the coherent states of the $SU(1,1)$ group 
in different space representations: Hilbert space and unit disk space or Lobachevian space \cite{Perel86, Klaud85}. We will briefly review geometrical aspects of coherent states 
and generalized coherent ones based on Lie algebras, and especially the ${\mathfrak {su}}(1,1)$ Lie algebra \cite{Fulton1991, Brif1997, Fujii2002, Gilmore2006, Gilmore2008}. The 
topological space that parametrizes the elements of a Lie group is a manifold. Therefore a Lie group is a finite dimensional manifold.  
  
Generally, group methods are used in order to find degenerate states of a given state in quantum mechanics \cite{Sazonova2000, Tinkham2003}. The Lie group of $SU(1,1)$ is widely 
used in physics as it plays the role of a spectrum generating dynamical group for a large number of dynamical systems, such as the Hydrogen atom, the harmonic oscillator or 
many-body systems. The Lie algebra of the generalized special unitary group $SU(1,1)$ is widely used in quantum optics. The $SU(1,1)$ coherent states arise in the case of 
generation of boson pairs with zero spins in homogeneous alternating electric field or in the gravitation field of the expanding Universe \cite{Perel73}. In particular, W\'odkiewicz 
and Eberly \cite{Wodkiewicz1985} have discussed the role of the generalized coherent states of Perelomov \cite{Perel72, Perel86} associated with the Lie algebra of the $SU(1,1)$ 
group in connection with variance reduction. Single- and two-mode bosonic realizations of the ${\mathfrak {su}}(1,1)$ Lie algebra are tightly connected with nonclassical squeezed 
states of light \cite{Lo1993}. Coherence-preserving Hamiltonians associated with the $SU(1,1)$ generalized coherent states have been studied by Gerry \cite{Gerry1987, Buzek1990}. 
It should be emphasized that the $SU(1,1)$ generalized coherent states in \cite{Perel72} and \cite{Wodkiewicz1985} are special cases of the two-photon coherent states discussed by 
Yuen \cite{Yuen1976}. This fact has been used by Gerry \cite{Gerry1987, Buzek1990} who applied the $SU(1,1)$ formulation of the two-photon coherent states to the issue of the 
interaction of squeezed light with a nonlinear non-absorbing medium modelled as an anharmonic oscillator. The coherent states have been especially useful in quantum optics and 
quantum information processing (QIP) \cite{Gazeau2009, Fox2006, Haroche2006}. Each mode of the electromagnetic field may be formally described as a harmonic oscillator, and 
different quantum states of the oscillator correspond to different states of the field. The field from a single-mode laser operating far enough above threshold can be described for 
many purposes as a coherent state; it differs from a coherent state in that its phase drifts randomly. Coherent states also play an important role in mathematical physics 
\cite{Gazeau2009, Perel86, Klaud85, Gilmore2008}. The paper investigates the bosonic realization of the Lie algebra for the $SU(1,1)$ group, for (generalized) coherent 
states in the Fock space, in case of an ion confined within a nonlinear Paul trap. 

The $SU(1,1)$ group \cite{Gazeau2009, Combescure2012} consists of unimodular matrices of $2$ x $2$ dimension with unit determinant. It is also isomorphic to the following noncompact groups 
\cite{Wybourne1974}:
\begin{equation}
 SO(2, 1)\approx SU(1,1)\approx SL(2, {\mathbb R})\approx {\mathcal Sp}(2, {\mathbb R}) .
\end{equation}

For example, we can construct the three-dimensional noncompact group $SU(1,1)$ by taking the intersection of the unitary group $U(1,1)$ with $SL(2; \mathbb{C})$. All usual groups 
of linear algebra such as $GL(n, {\mathbb R})$, $SL(n, {\mathbb R})$, $O(n, {\mathbb R})$, $U(n)$, $SO(n, {\mathbb R})$, $SU(n)$, $Sp(n, {\mathbb R})$ are real or complex Lie 
groups \cite{Kirillov2008}.

\section{The Variational Principle}
\label{TDVP}
\subsection{The Variational Principle. Test states}\label{VPTS}

The Time-Dependent Hartree-Fock (TDHF) method was first proposed by Dirac in 1930. In this approximation (which is equivalent to the Hartree-Fock approximation) the total 
electronic wave function can be approximated by a product of one-electron wave functions. Furthermore, one must assume that the potential experienced by a given electron
is an average of the potentials produced by the remaining electrons. The reference state for TDHF methods is the single determinant Hartree-Fock ground state, represented as 
a Slater determinant (using second quantization) \cite{Linderberg2004}. 

The equations of motion in classical mechanics arise as solutions of variational problems. An example would be Fermat's principle of least time, stating that light propagates 
between two points by a path which takes the shortest possible amount of time. In a manner similar to light, all systems possessing only kinetic energy move along geodesics, which 
are paths that minimize energy. As a mechanical system usually possesses both kinetic and potential energy, the quantity to be minimized is the mean value of the 
difference between the kinetic and potential energy. This less intuitive quantity is called the action \cite{McDuff98}. The variational principle results from the definition of an 
action integral $S=\int\limits_{t_1}^{t_2}L\left( \psi \right) dt\ $, where the real Lagrange function of this variational problem is defined as \cite{Krame81, Jost2009} 
\begin{equation}
\label{v2}L\left( \psi \right) =\frac{1}{\left\langle \psi |\psi \right\rangle }{\left[\left\langle \psi |H|\psi 
\right\rangle - \Im{m}\left\langle \frac{\partial \psi} {\partial t}\Bigg|\psi \right\rangle \right]} , 
\end{equation}
where $\hbar H$ is the quantum Hamiltonian of the system. In eq. (\ref{v2}) $\psi $ is a vector in the Hilbert space ${\mathcal H}$ for any moment of time $t$, and it belongs to 
the domain of the self-adjoint quasienergy operator $K(t)=H-i\partial/\partial t$. The variational principle for $S$ applied on the entire Hilbert space ${\mathcal H}$, leads to 
the Schr\"odinger equation \cite{Krame81, Drago04}. 

By minimizing the action $\left(\delta S=0\right)$, it can be shown that the Schr\"odinger equation is rigurously obtained from the variational principle. The natural symplectic 
structure induced by the transition probability between states for unitary transformations, achieved by the identification of the symplectic form with the imaginary part of the 
scalar product, enables approaching quantum mechanics issues through the formalism of the classical mechanics on (infinitely dimensional) symplectic manifolds (phase spaces), in 
particular K\"ahler manifolds. 

We apply the variational principle to a manifold ${\widehat {\mathcal M}}$ of test vectors, parametrized by the points of a finite $2n$ dimensional phase space ${\mathcal M}$, with 
a K\"ahler manifold structure. In agreement with the Darboux theorem, the symplectic manifold ${\widehat {\mathcal M}}$ admits canonical local coordinates and associated complex 
local coordinates. Generally the complex structure is not global, but local complex parametrization is preferred. In case of elementary quantum systems with dynamical symmetry 
groups which admit coherent states, the complex structure is global and ${\mathcal M}$ represents a K\"ahler manifold. 

\subsection{The Lagrange function. Hamilton equations of motion}\label{Lagrange}

We will consider $z=\left( z_1,z_2,\ldots ,z_n\right) \in {\mathcal O}$ as a system of complex canonical local coordinates in ${\mathcal M}$, where ${\mathcal O}$ is an open set 
from ${\mathbb C}^n$. ${\mathcal M}$ is a manifold of dimension $2n$. The (global) symplectic structure and the (local) K\"ahler structure are induced on the variety of 
test vectors ${\widehat {\mathcal M}}$. We choose a family of vectors $\psi \left( z\right) $ $\in \widehat{{\mathcal M}}$ with $z\in {\mathcal O}$, holomorphic in $z$, such as 
${\partial \psi \left( z\right) }/{\partial z^*_i}=0 \,,\ i=1,\ldots ,n\ $. The Lagrange function in the complex parametrization can be expressed as 
\begin{eqnarray}
\label{v3}L\left( z,\, z^*\right) =\frac i{2\left\langle \psi \left( z^*\right)
|\psi \left( z\right) \right\rangle }\sum_{i=1}^n\left\{ \dot
z_j\left\langle \psi \left( z^*\right) \Bigg|\frac{\partial \psi \left(
z\right) }{\partial z_i} \right\rangle -{\dot{z}^*_j}\left\langle \frac{\partial \psi \left( z^*\right) }
{\partial z_j}\Bigg|\psi \left( z\right) \right\rangle \right\}- \nonumber \\
-\frac{\left\langle \psi \left( z^*\right) |H|\psi \left(z\right) \right\rangle }{\left\langle \psi \left( z^*\right) |
\psi \left(z\right) \right\rangle }\, .
\end{eqnarray}
It is convenient to introduce the following notations \cite{Krame81}:
\begin{equation}
\label{v19}N\left( z,z^*\right) =\left\langle \psi \left( z^*\right)
|\psi \left( z\right) \right\rangle \;,\;\;H_{cl}\left( z.z^*\right) =%
\frac{\left\langle \psi \left( z^*\right) |H|\psi \left( z\right)
\right\rangle }{\left\langle \psi \left( z^*\right) |\psi \left( z\right)
\right\rangle }\ , 
\end{equation} 
where $H$ is the quantum Hamiltonian of the system, while $H_{cl}$ stands for the classical Hamiltonian. Then the Lagrange function described by eq. (\ref{v3}) becomes
\begin{equation}
\label{v20}L\left( z,\ z^*\right) =\frac i2\left( \dot z\nabla _z-{\dot{z^*}}\nabla _{z^*}\right) \ln N\left
( z,z^*\right) -H_{cl}\left( z,z^*\right) \,, 
\end{equation}
where $z\cdot \nabla _z=\sum\limits_{i=1}^nz_i\partial _{z_i}$. Hence the variation of the action is
\begin{equation}
\label{v28}\delta S=\int \left\{ i\sum_{j,k=1}^n\frac{\partial ^2\ln N}{%
\partial z_j\partial z^*_k}\ \dot z_j\delta z^*_k-i\sum_{j,k=1}^n\frac{%
\partial ^2\ln N}{\partial z^*_j\partial z_k}\ {\dot z^*}_j\delta z_k-\delta H_{cl}\right\} \ . 
\end{equation}
From the condition of extreme $\delta S=0$, using independent variables, we obtain a system of equations : 
\begin{equation}
\label{v31}i\sum_{j=1}^n\frac{\partial ^2\ln N}{\partial z^*_j\partial z_k}{\dot z}^*_j=-\frac{\partial H_{cl}}
{\partial z_k}\;,\;\;i\sum_{j=1}^n\frac{\partial ^2\ln N}{\partial z_j\partial z^*_k}\dot z_j=-\frac{%
\partial H_{cl}}{\partial z^*_k}\;. 
\end{equation}   
We now introduce the matrix of the symplectic structure on $\widehat{\mathcal M}$
\begin{equation}
\label{v29}\Omega =\left( \omega _{jk}\right) _{1\leq j,\,k\leq n}\ ,\;\
\omega _{jk}=\frac{\partial ^2\ln N}{\partial z_j\partial z^*_k}\;\;, 
\end{equation}
and Eqs. (\ref{v31}) change accordingly into
\begin{equation}
\label{v32}i\sum_{j=1}^n\omega _{kj}{\dot z}^*_j=-\frac{\partial H_{cl}}{%
\partial z_k}\;,\;\;i\sum_{j=1}^n\omega _{jk}\dot z_j=\frac{\partial H_{cl}}{%
\partial z^*_k} \,,
\end{equation}
where $\omega^*_{jk}=\omega _{kj}$. Hence we can ascertain that the matrix $\Omega =\omega _{ij}$ is Hermitian. The Poisson brackets for the 
$f,\,g\in C^\infty \left( {\mathcal M}\right)$ functions, smooth on $\mathcal M$, can be introduced as:
\begin{eqnarray}
\label{v33}\left\{ f,g\right\} =i\sum_{j,k=1}^n\left( \lambda _{jk}\frac{\partial f}{%
\partial z_j}\frac{\partial g}{\partial z^{*}_k}-\lambda^*_{kj}\frac{%
\partial g}{\partial z_j}\frac{\partial f}{\partial z^*_k}\right) = \nonumber \\
=\left(\frac{\partial f}{\partial z}\right)^t\Omega^{*-1}\,\frac{\partial g}{\partial z^*}+
\left(\frac{\partial f}{\partial z}\right)^t\Omega^{-1}\,\frac{\partial g}{\partial z} \,,
\end{eqnarray}
where $\lambda_{jk}=-i{\left(\Omega^*\right)}^{-1}$. Thus the variational principle for $S$ applied on ${\widehat {\mathcal M}}$ leads to the following classical Liouville 
equations of motion \cite{Krame81}:
\begin{equation}
\label{v34}\frac{dz_j}{dt}=\left\{z, H_{cl} \right\} \, \   ,\,\, \frac{dz^*_j}{dt}=\left\{ z^*,H_{cl}\right\} \, \   ,   
\end{equation}
where
\begin{equation}\label{v35}
H_{cl}\left(z,z^*\right)=\frac{\left\langle\psi \left(z^*\right)|H|\psi\left(z\right)\right\rangle}
{\left\langle \psi\left(z^*\right)|\psi \left(z\right)\right\rangle} .
\end{equation}

This paper deals with the study of classical and semiclassical nonlinear dynamical systems. It focuses on similarities and differences found when comparing the dynamical behaviour 
of Hamiltonian systems described by classical theory in respect with those described by the quantum theory \cite{Plimak1994, Polko2009}. An algorithm results, which associates 
an energy function to the quantum Hamiltonian. This function represents a classical type Hamiltonian, whose values are precisely the expected values of the quantum
Hamiltonian \cite{Gazeau2009} on coherent states. The energy function determines nonlinear equations of motion which can be approached using the theory of differential dynamical 
systems. We suggest that such algorithm could be used in order to establish a correspondence between semiclassical and classical dynamics.

The phase space ${\mathcal M}$ is a symplectic manifold described by eq. (\ref{v29}), with Poisson brackets determined by eq. (\ref{v34}). Eqs. (\ref{v31}) are the Hamilton 
equations which describe the Hamiltonian function $H_{cl}$ on ${\mathcal M}$. Thus $H_{cl}\left(z,z^*\right)$ is the expected value of the quantum Hamiltonian in the state 
represented by $\psi (z) \in \widehat {\mathcal M}$. Consequently $H_{cl}$ is considered the classical Hamiltonian associated to the quantum Hamiltonian $H$. This association is 
usually called {\it {dequantization}} \cite{Abrikosov2005} and it represents a set of rules which turn quantum mechanics into classical. The solutions of the equations of motion 
for the classical Hamiltonian $H_{cl}$ contain the same amount of spectral information as the solutions of the Schr\"odinger equation for the quantum Hamiltonian $H$, in case of 
many of the dynamical systems of interest. One of the conditions that are to be satisfied is to choose $\widehat{\mathcal M}$ as a manifold of coherent states.   

The group of linear canonical transformations of a dynamical system with $n$ degrees of freedom is the symplectic group ${\mathcal Sp}\left(2n, {\mathbb R}\right)$. An electrically 
charged particle confined within an electromagnetic trap can be explicitly described using coherent states, for a subgroup ${\mathcal G}$ of the symplectic group 
${\mathcal Sp}\left(6, {\mathbb R}\right)$ \cite{Major2005, Werth2009}. Symplectic transformations in ${\mathcal Sp}\left(2n, {\mathbb R}\right)$ leave invariant the canonical form 
of the classical Hamiltonian equations of motion \cite{McDuff98, Kirillov2008}.  

\subsection{Symplectic coherent states. Algebraic models. Equations of motion}
\label{Symplectic}

Further on we will introduce the coherent states built over weight states of the discrete positive series of the symplectic group ${\mathcal Sp}\left( 2,{\mathbb R}\right)$, which 
is also a real Lie group \cite{Major2005, Combescure2012, Gheor92, Gheor2000, Mihalcea2006}. We will consider a unitary representation ${\mathcal U}$ of the group ${\mathcal Sp}\left( 2,{\mathbb R}\right) $ 
in the Fock-Hilbert space ${\mathcal H}$. We will denote by ${\mathcal G}$ the group of unitary operators ${\mathcal U}\left( g\right) $ with $g\in {\mathcal Sp}\left( 2,{\mathbb R}\right) $ 
and by ${\mathfrak g}$ the Lie algebra for this group (in fact, the $SU(1,1)$ group Lie algebra, denoted by ${\mathfrak {su}}(1,1)$). The ${\mathfrak {su}}(1,1)$ algebra is 
determined by the three generators $K_0,K_1,$ and $K_2$, introduced in eqs. \ref{dinq1}. Further on we recall the ladder operators defined in eq. \ref{dinq2}, such as the Lie 
algebra may be cast into canonical form \cite{Lo1993, Brif1997, Dodon2003}, as described by eq. \ref{dinq1}. The Lie algebra ${\mathfrak {sp}}\left( 2,{\mathbb R}\right) $ consists 
of all the real linear combinations of the operators $iK_0, iK_1$, and $iK_2$. We will denote by ${\mathcal U}_k$ a unitary irreducible subrepresentation of the ${\mathcal U}$ 
representation in the Hilbert subspace ${\mathcal H}_k$ of the ${\mathcal H}$ space, where $k$ stands for the Bargmann index \cite{Bargmann1947}. The discrete series unitary 
representation of the Lie algebra considered are labelled by the eigenstates of the Casimir operator C: 
\cite{Combescure2012, Gheor2000}
\begin{equation}
\label{dinq72}C_2=K_0^2-K_1^2-K_2^2 \;,
\end{equation}
with eigenvalues $k\left( k-1\right)$ \cite{Sazonova2000, Dragt86, Popov2005}. We will consider only unitary irreducible representations (UIR) of the  group 
${\mathcal Sp}\left( 2,{\mathbb R}\right) $, corresponding to the discrete  positive series ${\mathcal D}^{+}\left(k\right) $ \cite{Dodon2003, Gheor2000, Sazonova2000, Lo1993}. 

Coherent and squeezed states in Quantum Optics are concerned with the Positive Discrete Series $D^+(k)$. The canonical ortonormal basis of the ${\mathcal H}_k$ space consists of the 
$\left| m,k\right\rangle ;\ \ m=0,1,\ldots ,$ vectors. For the representations of interest, the states $\left|m, k\right\rangle$ diagonalize the compact operator $K_0$. The operators 
$K_+$ and $K_-$ are Hermitian conjugates of each other and act as raising and lowering operators of the quantum number m \cite{Dodon2003, Gheor92, Fujii2002}: 
\begin{equation}
\label{dinq73}
\begin{array}{c}
K_{+}\left| m,k\right\rangle =\left[ \left( m+1\right) \left( m+2k\right)\right] ^{1/2}\left| m+1,k\right\rangle \;, \\ 
K_{-}\left| m,k\right\rangle =\left[ m\left( m+2k-1\right) \right]^{1/2}\left| m-1,k\right\rangle \;, \\ 
K_0\left| m,k\right\rangle =\left( m+k\right) \left| m,k\right\rangle \ , k > 0 , \ m = 0, 1, 2, \ldots . 
\end{array}
\end{equation}
The coherent states for the symplectic group ${\mathcal Sp}\left( 2,{\mathbb R}\right) $ are defined as \cite{Dodon2003}: 
\begin{equation}
\label{dinq74}\left| z,m,k\right\rangle =U\left( z\right) \left| m,k\right\rangle \ , 
\end{equation}
Using the disentagling theorem for the $SU(1,1)$ Lie algebra \cite{Wodkiewicz1985, Dattoli1988}, the unitary operator $U(z)$ can be expressed as  
\begin{equation}
\label{dinq75}U\left( z\right) =\exp \left( zK_{+}\right) \exp \left( \beta K_0\right) \exp \left( -\bar zK_{-}\right) \ , 
\end{equation}
where $\left| z\right| <1$ and $\beta =\ln \left( 1-z\bar z\right)$. The $\left| z,m,k\right\rangle $ states are called {\it symplectic coherent states}. The phase space 
${\mathcal M}$ whose points parametrize the coherent states is the unitary disk $\left| z\right| <1$, endowed with the Lobacevski metrics $ds^2=4\left( 1-z\bar z\right) ^{-2}dz\,d\bar z$. 
For $m=0$ we infer the geometrical construction of Perelomov \cite{Perel86, Lo1993}. Eq. (\ref{dinq74}) remains valid for coherent states with $m>0$, realized by applying the 
unitary operators $U\left( z\right) \in {\mathcal G}$ on the nondominant weighting vectors $\left| k,m\right\rangle ;\ m=1,2,\ldots \ $. 

Using the Baker-Campbell-Hausdorff formula \cite{Feng1980, Kirillov2008, Sternberg2004, Combescure2012} and eq. (\ref{dinq74}) we infer 
\cite{Gheor92, Gheor2000, Mihalcea2011}:
\begin{equation}
\eqalign{U^{-1}\left( z\right) K_0U\left( z\right) =\left( 1-z\bar z\right)^{-1}\left[ \bar zK_{-}+\left( 1+z\bar z\right) K_0+zK_{+}\right] \ , \cr 
U^{-1}\left( z\right) K_{+}U\left( z\right) =\left( 1-z\bar z\right)^{-1}\left[ K_{+}+2\bar zK_0+\bar z^2K_{-}\right] \ , \cr
U^{-1}\left( z\right) K_{-}U\left( z\right) =\left( 1-z\bar z\right)^{-1}\left[ K_{-}+2zK_0+z^2K_{+}\right] \ .}\label{dinq77}  
\end{equation}
We will denote by $\Omega _\varepsilon =K_0+\varepsilon K_1$, where $\varepsilon =\pm 1$. Then
\begin{equation}
\label{dinq79}U^{-1}\left( z\right) \Omega _\varepsilon U\left( z\right) =\frac{\left( 1+\varepsilon z\right) \left( 1+\varepsilon \bar z\right) }{1-z\bar z}E_\varepsilon \ , 
\end{equation}
\begin{equation}
\label{dinq80}E_\varepsilon =\frac 12\left( 1-z\bar z\right) ^{-1}\left[2\left( 1+z\right) \left( 1+\bar z\right) K_0+\varepsilon \left( 1+\bar z\right) ^2K_{-}+\varepsilon \left( 1+z\right) ^2K_{+}\right] 
\end{equation}
From eqs. (\ref{dinq77}) - (\ref{dinq80}), we infer 
\begin{equation}
\label{dinq93}\left\langle z,k,m|\Omega _\varepsilon ^n|z,k,m\right\rangle =\left[ \frac{\left( 1+\varepsilon z\right) \left( 1+\varepsilon \bar z\right) }{1-z\bar z}\right] ^n\ 
\left\langle k,m|E_\varepsilon^n|k,m\right\rangle \ . 
\end{equation}
Eq. (\ref{dinq93}) enables the explicit calculus of the Husimi function for every algebraic model with a dynamical group ${\mathcal G}$. If $H$ is the Hamiltonian for such a model, 
expressed as a polynomial in the operators $K_0,K_1,$ and $K_2$, we observe that the Hamiltonian function $H_{cl}\left(z\right) {=}\left\langle z,k,m|H|z,k,m\right\rangle $ leads 
to the following equation of motion in the phase space ${\mathcal M}$ : 
\begin{equation}
\label{dinq106a}\dot z=\left\{ z,H_{cl}\right\} \ .
\end{equation}
The brackets $\left\{ \ ,\ \right\} $ stand for the Poisson brackets associated with the symplectic manifold ${\mathcal M}$, defined as \cite{Krame81, Drago04} 
\begin{equation}
\label{dinq107a}\left\{ f,g\right\} =\frac{\left( 1-z\bar z\right) ^2}{2i\left( k+m\right) }\left( \frac{\partial f}{\partial z}\frac{\partial g}{\partial \bar z}-\frac{\partial f}
{\partial \bar z}\frac{\partial g}{\partial z}\right) \ , 
\end{equation}
where the phase space classical observables $f$ and $g$ are smooth and real functions of $\Re e\,z$ and $\Im m\,z$. When $m=0$, from the TDVP applied to the coherent states 
$\left| z,\,0,\,k\right\rangle $ for the Hamiltonian $H$ we infer Eqs. (\ref{dinq106a}), which can also be expressed as 
\begin{equation}
\label{dinq108a}\dot z=\frac{\left( 1-z\bar z\right) ^2}{2ik}\frac{\partial H_{cl}}{\partial \bar z}\ . 
\end{equation}
We now introduce the complex variables $\xi $ and $\eta $ defined as 
\begin{equation}
\label{dinq102a}\xi =\frac{\left( 1+z\right) \left( 1+\bar z\right) }{1-z\bar z}\;\ \ ,\;\;\ \eta =\frac{\left( 1-z\right) \left( 1-\bar z\right) }{1-z\bar z}\;\ . 
\end{equation}
Then eqs. (\ref{dinq108a}) become
\begin{equation}
\label{dinq124}\dot \xi =\frac 2{ik}\frac{z-\bar z}{1-z\bar z}\frac{\partial H_{cl}}{\partial \eta }\ ,\;\dot \eta =-\frac 2{ik}\frac{z-\bar z}{1-z\bar z}\frac{\partial H_{cl}}{\partial \xi }\ ,  
\end{equation}
which are exactly the equations of motion in the phase space for the system we considered. The group of the linear canonical transformations of a dynamical system system with $n$ 
degrees of freedom, is the symplectic group ${\mathcal Sp}\left(2n, \mathbb R \right)$ \cite{Wybourne1974}. In case of dynamical symmetry groups, the Schr\"odinger equation solutions 
associated to linear Hamiltonians are given by coherent vectors multiplied by geometrical (Berry) phase factors \cite{Gheor92}. The associated classical Hamiltonians result from the 
expected values  of the quantum Hamiltonians on coherent symplectic states. Thus spectral information is coded into the phase portrait. The formalism presented here allows to 
explicitly construct bases and systems of symplectic coherent states for the study of trapped ion systems.

\section{Semiclassical dynamics of an ion in combined traps with axial symmetry}\label{Quadyn}

\subsection{Algebraic models for nonlinear axial traps}\label{AlgebraicModels}

We have reviewed the properties of the coherent states and their exceptional importance for physics and especially for quantum optics. Coherent states for charged particles 
confined in electrodynamic traps can be built using the group theory \cite{Combescure2012, Combe87, Nieto2000}. We will further investigate the semiclassical dynamics of an ion trapped in a 
nonlinear electrodynamic trap with cylindrical symmetry, using the symplectic state approach and coherent state formalism developed in \cite{Perel86, Major2005, Werth2009, Gheor92}, 
presented in Sections \ref{Lagrange} and \ref{Symplectic}. The coherent state formalism is applied in order to study the semiclassical behaviour of trapped ions 
\cite{Major2005, Werth2009, Gheor92, Gheor2000}. We will introduce the Lie algebra generators of the axial symplectic group ${\mathcal Sp}\left(2,{\mathbb R}\right)_a$
\cite{Major2005, Gheor92}
\begin{equation}\label{quadyn1}
K_{0,1a}=\frac{M\omega _a}{4\hbar }\left[ z^2\pm\frac{p_z^2}{M^2\omega _a^2}\right] , \;
K_{2a}=\frac i{4\hbar }\left[ 2z\; \frac \partial {\partial z}+1\right]\;,
\end{equation}
and the Lie algebra generators of the radial symplectic group ${\mathcal Sp}\left( 2,{\mathbb R}\right)_r$ for a fixed eigenvalue of the orbital angular momentum $\hbar l$ 
\begin{equation}
\eqalign{K_{0,1r}=\pm\frac 1{2\hbar \omega _r}\left[ \frac 1{2M}\ p_\rho ^2 \pm \frac{M\omega _r^2}2\rho ^2+\frac{\hbar ^2}{2M}\left( l^2-\frac 14\right) \frac 1{\rho ^2}\right]\;,\cr 
K_{2r}=\frac i{4\hbar }\left[ 2\rho \;\frac \partial {\partial \rho}+1\right] \;.}\label{quadyn2}
\end{equation}
We have denoted by $\omega _a/2\pi $ the axial frequency and by $\omega _r/2\pi $ the radial frequency respectively, of the quadrupole trap potential. The $K_{0j}, K_{1j}$ and 
$K_{2j}=i[K_{1j}, K_{0j}], j=a,r $ operators satisfy the commutation relations for the Lie algebra of the symplectic group ${\mathcal Sp}\left(2,{\mathbb R}\right)$
\begin{equation} \label{quadyn3}
[K_{0j}, K_{1j}]=iK_{2j}, \ \ [K_{2j},K_{0j}]=iK_{1j}, \ \ [K_{2j}, K_{1j}]=iK_{0j} .
\end{equation}

Then the quantum Hamiltonian of a particle with electrical charge $Q$, mass $M$ and 
orbital angular momentum $\hbar l$, located in the quadrupole trap with octupole anharmonicity, can be expressed as 
\cite{Mihalcea2011}
\begin{eqnarray}H_l=-\frac{4Q\hbar }{M\omega _a}A\left( t\right) \left( K_{0a}+K_{1a}\right) + \frac \hbar {\omega _r}\left( \frac{\omega _c^2}4+\frac{2Q}M A\left( t\right) \right) 
\left( K_{0r}+K_{1r}\right) - \nonumber \\
\frac{\omega _c}2\hbar l +\hbar \omega _a\left( K_{0a}-K_{1a}\right) +\hbar \omega_r\left( K_{0r}-K_{1r}\right) +c_2A\left( t\right) H_4\left( \rho ,z\right) \ ,\label{quadyn4}
\end{eqnarray}
where $\omega _c$ stands for the cyclotronic frequency and $H_4$ is a harmonic polynomial of degree 2 homogeneous in the variables $z^2$ and $\rho ^2=x^2+y^2$. These variables can 
be expressed as a function of the generators 
\begin{equation}\label{quadyn5}
z^2=\frac{4\hbar }{M\omega _a}\left( K_{0a}+K_{1a}\right) \ ,\;\ \ \rho ^2=\frac{4\hbar }{M\omega _r}\left( K_{0r}+K_{1r}\right) \ .
\end{equation}
The coupling constant $c_2$ and the time dependent function $A\left( t\right) $ are characteristic for the type of trap used and will be explained later. 

Eq. (\ref{quadyn4}) results from the general expression of the Hamiltonian associated to a particle which undergoes the influence of an electromagnetic field, characterized by an 
electrical potential $\Phi $ and a magnetic induction $\vec B$ \cite{Grynberg2010, Major2005, Werth2009, Mihalcea2006}:
\begin{equation}\label{quadyn6}
H=\frac 1{2M}\left[ \vec p-\frac 12\;Q\vec B\times \vec r\right]^2+Q \Phi \ .
\end{equation}
The momentum operator for the particle is $\vec p=-i\hbar \nabla $, while the magnetic field is considered to be axial: $\vec B=\left( 0,0,B_0\right) $ and $\omega_c=QB_0\backslash M$. 
We can introduce the (axial) angular momentum operator on the $z$ axis, defined as $L_z=xp_y-yp_x$. Then Eq. (\ref{quadyn6}) becomes
\begin{equation}\label{quadyn7}
H=\frac 1{2M}\left( p_x^2+p_y^2+p_z^2\right) +\frac 1{8M}\;Q^2B_0^2\;\rho^2 -\frac {QB_0}{2M}\;L_z + Q\Phi \  .
\end{equation}
Because the axial angular momentum operator $L_z$ commutes both with the symplectic groups generators as well as with the Hamiltonian $H$, the study of the quantum system can be 
restricted to a subspace of the Hilbert space for which $L_z=\hbar lI$, where $l$ is the orbital quantum number. The Hamiltonian reduced to this subspace can be expressed as 
\begin{equation}\label{quadyn8}
H_l=\frac 1{2M}\left( p_x^2+p_y^2+p_z^2\right) +\frac 1{8M}\;Q^2B_0^2\;\rho ^2- \frac{QB_0}{2M}\;\hbar l+Q\Phi \ .
\end{equation}
Choosing a system of cylindrical coordinates $\rho $ and $\theta $, the expressions for the radial coordinates are $x=\rho \cos \theta $ and $y=\rho \sin \theta $. It follows that 
\begin{equation}\label{quadyn9}
p_x^2+p_y^2=-\hbar ^2\left( \frac{\partial ^2}{\partial \rho ^2}+\frac 1{\rho ^2}\frac{\partial ^2}{\partial \theta ^2}+\frac 1\rho \frac \partial {\partial \rho }\right) \ ,\;\ 
L_z=-i\hbar \ \frac \partial {\partial \theta }\ \ .
\end{equation}
We choose $\Psi =e^{il\theta }\chi $ as a solution of the Schr\"odinger equation, where $\chi $ depends only on $\rho $ and $z$. Then $L_z\Psi =\hbar l\Psi $ and $\chi $ is a 
solution of the Schr\"odinger equation for the Hamiltonian
\begin{equation}\label{quadyn10}
H_l=-\frac{\hbar ^2}{2M}\left( \frac{\partial ^2}{\partial \rho ^2}-\frac{l^2}{\rho ^2}+\frac 1\rho \frac \partial {\partial \rho }\right) -\frac{\hbar ^2}{2M}\frac{\partial ^2}{\partial z^2}+\frac{M\omega _c^2}8\rho ^2-\frac{\omega _c}2\;\hbar l+Q\Phi \ ,
\end{equation}
which results from Eqs. (\ref{quadyn8}) and (\ref{quadyn9}).

We will now choose a potential with axial symmetry, such as
\begin{equation}
\label{quadyn11}\Phi \left( \vec r,t\right) =A\left( t\right) g\left( \rho , z\right) \ ,
\end{equation}
where $A\left( t\right) $ is a time-periodical function of period $T=2\pi /\Omega $, while $g$ is a function of $\rho ^2$ and $z^2$ which satisfies
\begin{equation}
\label{quadyn12}g\left( \rho ,z\right) =\sum_{k\geq 1}c_kH_{2k}\left( \rho , z\right) \ .
\end{equation}
The $H_{2k}$ polynomials are harmonic polynomials of degree $k$ in $\rho ^2$ and $z^2$. The potential described by Eq. (\ref{quadyn11}) is characteristic for a trap with axial 
symmetry and with symmetry in respect to the radial plane $xOy$. For a Penning trap $A$ is constant in time, but for most of the radiofrequency traps we choose 
$A\left( t\right) =U_0+V_0\cos \Omega t$. In case of a harmonic potential $\left( c_k=0\ \textrm{for}\ k>1\right) $ 
we have
\begin{equation}
\label{quadyn13}g\left( \rho ,z\right) =\frac 1{r_0^2+2z_0^2}\left( \rho ^2-2z^2\right) ,\;\ c_2=-\frac 
1{r_0^2+2z_0^2}\;,
\end{equation}
where $r_0$ and $z_0$ are the semiaxes of the combined quadrupole trap we have considered. We remind that a combined trap consists of a Penning trap and a superimposed Paul trap. 
The particular case of the ideal Paul trap is obtained for $\vec B=0$. The quadrupole potential is produced by hyperbolic electrodes. Real traps are different from ideal ones. 
Due to mechanical imperfections (holes drilled in the endcap electrodes or in the ring electrode, the latter causing a radial asymmetry of the potential), the trap potential is not 
a pure quadrupole potential. The potential is then expanded in a power series and higher order terms have to be considered. Due to the trap symmetry we have chosen, the odd terms 
of higher order from the series expansion of the potential vanish and we are left only with the even terms. In case of a quadrupole trap with octupole anharmonicity, the electric 
potential can be expressed as 
\begin{equation}
\label{quadyn14}g\left( \rho , z\right) =c_1H_2\left( \rho , z\right)+c_2H_4\left( \rho , z\right) \ ,
\end{equation}
\begin{equation}
\label{quadyn15}H_2\left( \rho ,z\right) =2z^2-\rho ^2\;,\;\ H_4\left( \rho
,z\right) =8z^4-24z^2\rho ^2+3\rho ^4\;.
\end{equation}
where $H_2$ is the pure quadrupole term and $H_4$ is the octupole term. Under such case, the Hamiltonian $H_l$ is 
given by Eq. (\ref{quadyn4}). The next anharmonical contribution is given by the polynomial 
\begin{equation}
\label{quadyn16}H_6=16z^6-120z^4\rho ^2+90z^2\rho ^4-5\rho ^6\ .
\end{equation}

The sketches for contours of the electrical potentials in the quadrupole and octupole trap, as well as in the case of the combined $H2+0.2 H4$ 
trap are given below

\begin{figure}
\begin{center}
\includegraphics{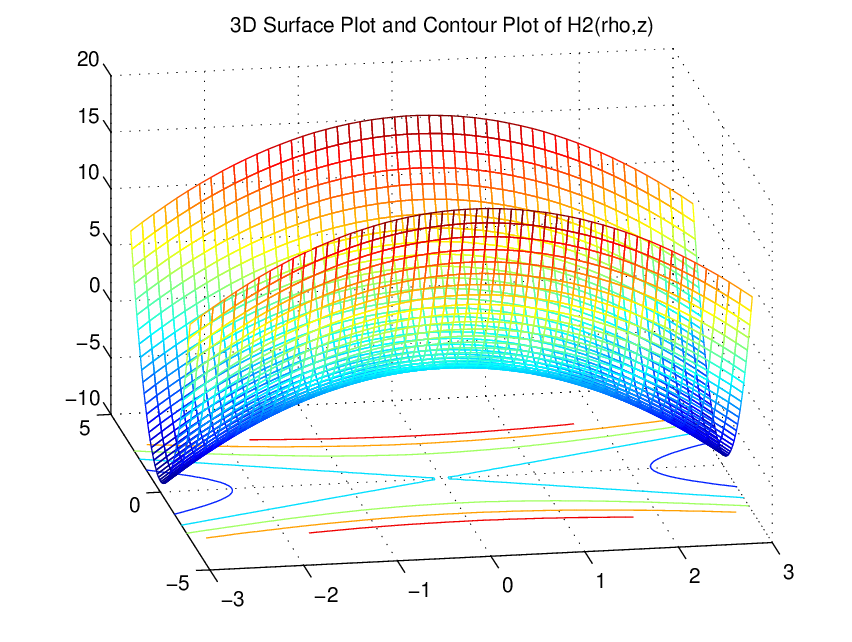} 
\end{center}

\begin{center}
\includegraphics{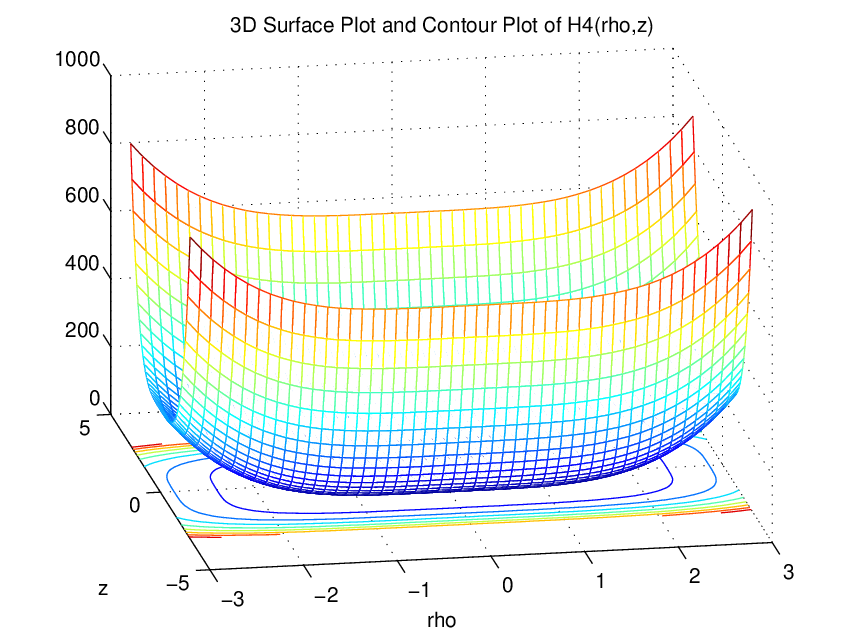} 
\end{center}


\caption{Surface plot and contour plot of the electric potential for the quadrupole $H2(\rho, z)$ and octupole $H4(\rho, z)$ traps}
\label{electricpotentials}
\end{figure}

By denoting
$$
K_r=\frac{M\omega _c^2}4-2Qc_2A\left( t\right) \;,\;\;K_a=4Qc_2A\left(t\right) \;,
$$
the Hamiltonian given by Eq. (\ref{quadyn10}) can be expressed as
\begin{eqnarray}\label{quadyn17}
H_l=-\frac{\hbar ^2}{2M}\left( \frac{\partial ^2}{\partial \rho ^2}-\frac{l^2}{\rho ^2}+\frac 1\rho \frac \partial {\partial \rho }\right) -\frac{\hbar ^2%
}{2M}\frac{\partial ^2}{\partial z^2}-\frac{\omega _c}2\hbar l + \nonumber \\
+\frac{K_r}2\rho ^2+\frac{K_a}2z^2+QA\left( t\right) P\left(\rho ^2,z^2\right) \ ,
\end{eqnarray}
where the anharmonical part is defined as
\begin{equation}
\label{quadyn18}P\left( \rho ^2, z^2\right) =\sum\limits_{k\geq 2}c_kH_{2k}\left( \rho ,z\right) \ .
\end{equation}
In the following we will study the Schr\"odinger equation
$$
i\hbar \ \frac{\partial \chi }{\partial t}=H_l\,\chi\;,
$$
where $H_l$ is given by Eq. (\ref{quadyn17}), with particularizations for combined quadrupole and octupole traps. 
In agreement with Eqs. (\ref{quadyn4}), (\ref{quadyn5}) and (\ref{quadyn17}), The Hamiltonian $H_l$ describes an 
algebraic model when $P\left( \rho ^2,z^2\right) $ is a polynomial function. This model is linear for quadrupole 
traps ($P$ is a linear combination between $\rho ^2$ and $z^2$) \cite{Mihalcea2011}.

\subsection{Dynamical symmetries for an ion in an axial trap}\label{DynSym}

The algebraic model described by the Hamiltonian $H_l$ can be characterized by the dynamical symmetry group ${\mathcal G}={\mathcal G}_a\otimes {\mathcal G}_r$, realized as a 
direct product between the axial symplectic group ${\mathcal G}_a={\mathcal Sp}\left(2,{\mathbb R}\right) _a$ and the radial symplectic group ${\mathcal G}_r = 
{\mathcal Sp}\left(2,{\mathbb R}\right) _r$ \cite{Gheor2000, Gheor97}.

From Eqs. (\ref{quadyn2}) we obtain the commutation relations 
\begin{equation}
\label{quadyn19}\left[ K_{0a},K_{1a}\right] =iK_{2a}\ ,\;\ \left[K_{2a},K_{0a}\right] =iK_{1a}\ ,\;\left[ K_{1a},K_{2a}\right] =iK_{0a}\ ,
\end{equation}
for the Lie algebra ${\mathfrak g_{a}}$ of the group ${\mathcal G}_a$. The Casimir operator of the Lie algebra ${\mathfrak g_a}$ determines the Bargmann indexes $k$, and it can be 
expressed as \cite{Gheor92, Gheor2000, Fujii2002}
\begin{equation}
\label{quadyn20}C_{2a}=K_{0a}^2-K_{1a}^2-K_{2a}^2=-\frac 3{16}I=k\left(
k-1\right) I\ ,
\end{equation}
where $I$ is the unitary operator. From Eq. (\ref{quadyn20}) we infer the values of the two Bargmann indexes which characterize the axial motion
\begin{equation}
\label{quadyn21}k_{a+}=\frac 14\;\ ,\;\;k_{a-}=\frac 34\ .
\end{equation}
From Eqs. (\ref{quadyn2}) we obtain the commutation relations
\begin{equation}
\label{quadyn22}\left[ K_{0r},K_{1r}\right] =iK_{2r}\ ,\;\ \left[K_{2r},K_{0r}\right] =iK_{1r}\ ,\;\left[ K_{1r},K_{2r}\right] =iK_{0r}\ ,
\end{equation}
for the Lie algebra ${\mathfrak g_{r}}$ of the group ${\mathcal G}_r$. The Casimir operator of the Lie algebra ${\mathfrak g_r}$ is \cite{Dodon2003, Gheor2000, Mihalcea2011}
\begin{equation}
\label{quadyn23}C_{2r}=K_{0r}^2-K_{1r}^2-K_{2r}^2=\frac{l^2-1}4\ I=k\left(k-1\right) I\ .
\end{equation}
From Eq. (\ref{quadyn23}) we infer the Bargmann index $k_r=$ $\left( l+1\right) /2$, where $l$ stands for the quantum orbital number. Eq. (\ref{quadyn17}) can be now expressed as
\begin{eqnarray}\label{quadyn24}
H_l=\hbar \omega _r\left( K_{0r}-K_{1r}\right) +2\hbar \omega _a\left(K_{0a}-K_{1a}\right) -\frac{\omega _c}2\hbar l \nonumber \\
+\frac{K_r}2\frac{2\hbar }{M\omega _r}\left(K_{0r}+K_{1r}\right) +\frac{K_a}2\frac{2\hbar }{M\omega _a}\left(K_{0a}+K_{1a}\right) +QA\left( t\right) P\left( \rho ^2,z^2\right) \ .
\end{eqnarray}

Consequently, the study of the Hamiltonian for an ion confined in a quadrupole trap $\left( P=0\right) $ has been restricted to the study of a linear model for the dynamic group 
${\mathcal G}$, for which the unitary irreducible representations are characterized by the Bargmann indexes $k_a$ and $k_r$, fixed \cite{Gheor92}. The solutions of the Schr\"odinger 
equation for this elementary quantum system will result explicitly by using the coherent state formalism for the group ${\mathcal G}$. This formalism will be also used for algebraic 
models which describe nonlinear traps (with $P\neq 0$).

It can be shown that symplectic coherent states are realized in the trapped ion dynamics and a classical Hamiltonian (the Husimi or the energy function) can be associated to the 
quantum Hamiltonian. The Hamiltonian function leads to equations of motion in the classical Lobacevski phase space.  

\subsection{Symplectic coherent states representation}\label{SymplecRepr}

According to paragraph \ref{DynSym}, the Bargmann indexes for the axial motion are given by $k_{-}=\frac 14$ and $k_{+}=\frac 34$, while in case of the radial motion the expression 
is $k_l=\left( 1+l\right) /2$, where $l$ stands for the orbital quantum number. The canonical orthonormal basis of the Hilbert space for an irreducible unitary representation of 
the group ${\mathcal G}_j$, of Bargmann index $k_j$, with $j=a,\,r$, consists of the vectors $\phi _{jm}\ , \ \ m=0,1,\ldots $ and it is characterized by the equations
\begin{equation}
\label{dinq56}K_{0j}\phi _{jm}=\left( k_j+m\right) \phi _{jm},\quad
K_{-j}\phi _{jj}=0.
\end{equation}

We can introduce the coherent symplectic vectors $\psi _{jm}\left( z_j\right) $, defined as 
\begin{equation}
\label{dinq57}\psi _{jm}\left( z_j\right) =U_j\left( z\right) \phi _{jm},
\end{equation}
\begin{equation}
\label{dinq58}U_j\left( z_j\right) =\exp \left( z_jK_{+j}\right) \exp \left(
\beta _jK_{0j}\right) \exp \left( -\bar z_jK_{-j}\right) ,
\end{equation}
where $U_j\left( z_j\right) $ are unitary operators of the representation, $\beta _j=\ln \left( 1-z_j\bar z_j\right) $, and $z_j$ is a complex variable from the unit disk 
$\left| z_j\right| <1$. The phase space ${\mathcal M}$ associated to an ion confined within an axial symmetry trap represents the product of the axial and radial unit disks, 
realized as a bound domain in ${\mathbb C^2}$, made out of all points $\left( z_a, z_r\right) $ with $\left| z_a\right| <1$ and $\left| z_r\right| <1$. 
The coherent states for the dynamical group ${\mathcal G}$, parametrized with points from ${\mathcal M}$, can be expressed as
\begin{equation}
\label{dinq59}\Phi _{m_am_r}\left( z_a,\,z_r\right) =\psi _{am_a}\left(z_a\right) \psi _{rm_r}\left( z_r\right)\;.
\end{equation}

In case of quadrupole traps, the symplectic coherent states $\Phi _{m_am_r}\left( z_a,z_r\right) $ multiplied by Berry phase factors are solutions of the Schr\"odinger equation 
\cite{Gheor92}.

The energy function associated to the Hamiltonian (\ref{quadyn24}) represents a classical Hamiltonian $H_{cl}$ on ${\mathcal M}$, whose values are exactly the expected values of 
the Hamiltonian $H_l$ on coherent states $\Phi_{m_am_r}\left( z_a,z_r\right) $, such as
$$
{H}_{cl}=\hbar \omega _r\left( k_r+m_r\right) \frac{\left( 1-z_r\right)\left( 1-\bar z_r\right) }{1-z_r\bar z_r}+2\hbar \omega _a\left(k_a+m_a\right) \frac{\left( 1-z_a\right)%
\left( 1-\bar z_a\right) }{1-z_a\bar z_a}
$$
$$
+\frac{2\hbar K_r}{M\omega _r}\left( k_r+m_r\right) \frac{\left(1+z_r\right) \left( 1+\bar z_r\right) }{1-z_r\bar z_r}+\frac{2\hbar K_a}{%
M\omega _a}\left( k_a+m_a\right) \frac{\left( 1+z_a\right) \left( 1+\bar z_a\right) }{1-z_a\bar z_a}
$$
\begin{equation}
\label{dinq95}-\frac{\omega _c}2\hbar l+H_{anarm}\;,
\end{equation}
cu
\begin{equation}
\label{dinq96}H_{anarm}=QA\left( t\right) \left\langle P\left( \rho ^2,z^2\right) \right\rangle =QA\left( t\right) \sum_{k\geq 1}c_k\left\langle H_{2k}\left(\rho ,z\right)\right\rangle \ .
\end{equation}
We have denoted by $\left\langle X\right\rangle $ the expected value of the operator $X$ in the coherent state $\Phi _{m_am_r}\left( z_a,z_r\right) $. In particular we obtain
\begin{equation}
\label{dinq97}\left\langle H_4\right\rangle =8S_{2a}-24S_{1r}S_{2r}+3S_{2r}\ ,
\end{equation}

\begin{equation}
\label{dinq98}\left\langle H_6\right\rangle
=16S_{3a}-120S_{2a}S_{1r}+90S_{1a}S_{2r}-5S_{3r}\ .
\end{equation}
where 
$$
S_{jr}=S_j\left( z_r,k_r,m_r\right) \ \ ,\;\;S_{ja}=S_j\left(z_a,k_a,m_a\right) \ \ ,\;\;1\leq j\leq 3\ ,
$$
with
\begin{equation}
\label{dinq100}S_j\left( z,k,m\right) =\left[ \frac{\left( 1+z\right) \left(1+\bar z\right) }{1-z\bar z}\right] ^jQ_j\left( k,m\right) \;,
\end{equation}
and
\begin{equation}
\label{dinq101a}Q_1\left( k,m\right) = 2\left( k+m\right)
\end{equation}
\begin{equation}
\label{dinq101b}Q_2\left( k,m\right) = 2k\left( 2k+1\right) +12km+6m^2
\end{equation}
\begin{equation}
\label{dinq101c}Q_3\left( k,m\right) = 4k\left( k+1\right) \left( 2k+1\right) + 4mk\left( 5+12k\right) + 4m^2\left( 15k+1\right) + 20m^3
\end{equation}
The values of the indexes are :
$$
m_a,m_r=0,1,\ldots \;;\;k_a=\frac 14,\frac 34\ ;\;k_r=\frac{l+1}2\ .
$$
We will introduce the complex variables $\xi_j$ and $\ \eta_j,\;j=a,r$, defined as
\begin{equation}
\label{dinq102}\xi _j=\frac{\left( 1+z_j\right) \left( 1+\bar z_j\right) }{1-z_j\bar z_j}\;\ \ ,\;\;\ \eta _j=\frac{\left( 1-z_j\right) \left( 1-\bar z_j\right) }{1-z_j\bar z_j}\;\ .
\end{equation}
Then the classical Hamiltonian (the Husimi function) for an octupole trap in case of the pseudopotential approximation can be expressed as
$$
H_{cl}=A_r\eta _r+A_a\eta _a+B_r\xi _r+B_a\xi _a+\left( C_{20}\xi_r^2+C_{11}\xi _r\xi _a+C_{02}\xi _a^2\right)
$$
\begin{equation}
\label{dinq103}+\left( D_{30}\xi _r^3+D_{21}\xi _r^2\xi _a+D_{12}\xi _r\xi_a^2+D_{03}\xi _a^3\right) \ -\frac{\omega _c}2\hbar l.
\end{equation}
If $P$ is a polynomial of $\rho ^2$ and $z^2$, then $H_{cl}$ is a polynomial of $\xi_a$ and $\xi_r$. In order to find the points of minimum for the Husimi function, we have to 
solve the equations
\begin{equation}
\label{dinq103b}\frac{\partial H_{cl}}{\partial \xi _a}=0\;,\;\;\frac{\partial H_{cl}}{\partial \xi _r}=0\;,\;\;\frac{\partial H_{cl}}{\partial \eta _a}=0\;,\;\;\frac{\partial H_{cl}}{\partial \eta _r}=0\ .
\end{equation}

\begin{figure}
\begin{center}
\includegraphics{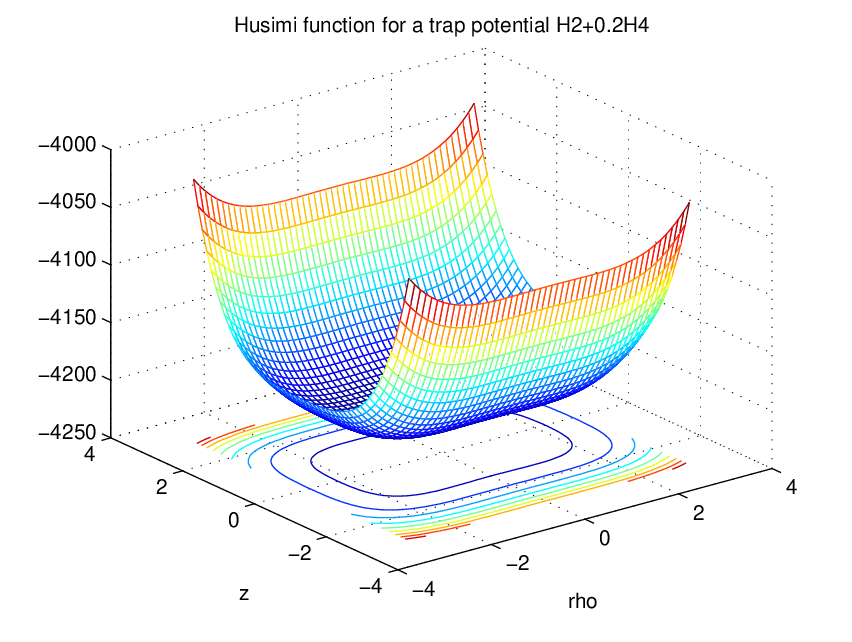} 
\end{center}

\begin{center}
\includegraphics{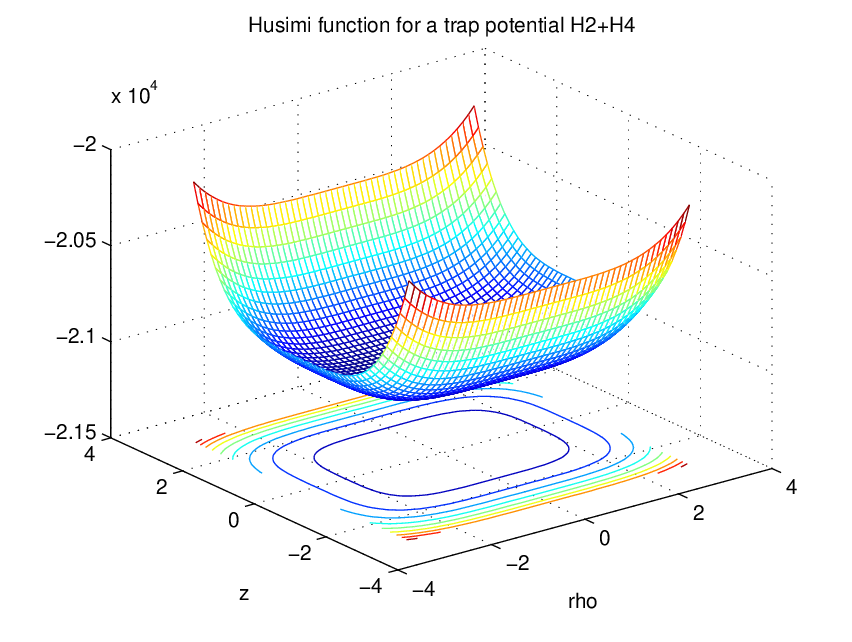} 
\end{center}


\caption{Husimi functions in case of fourth order traps with electric 
potentials $H2(\rho, z) + 0.2 H4(\rho,z)$ and $H2(\rho, z)+H4(\rho, z)$, respectively}
\label{HusimiFunctions}
\end{figure}

The Husimi functions for a fourth order trap are given in Fig. \ref{HusimiFunctions}.

It can be observed that the Hamiltonian function $H_{cl} = \left\langle z,k,m|H|z,k,m\right\rangle $ determines the classical Liouville equation of motion in the Lobacevski phase 
space $\left| z\right| <1:$%
\begin{equation}
\label{dinq106}\dot z=\left\{ z,H_{cl}\right\} \;,
\end{equation}
where $\left\{ \ ,\ \right\} $ stands for the generalized Poisson bracket, defined in Eq. (\ref{dinq107a}) \cite{Krame81}. Eq. \ref{dinq106} is exactly the standard form of the 
symplectic structure, as in classical mechanics \cite{Golds80, Feng1980}. After applying the TDVP to the coherent states $\left| z,\,0,\,k\right\rangle $ for the Hamiltonian $H$, 
we infer Eqs. 
\begin{equation}
\label{dinq107}
\left\{ z,H_{cl}\right\}=\frac{(1-z{\bar z})^2}{2 ik}\frac{\partial H_{cl}}{\partial \bar z} \,
\end{equation}
which represents a classical type equation of motion. If $P$ given by Eq. (\ref{quadyn18}) is a polynomial in $\rho^2$ and $z^2$, then $H_{cl}$ is a polynomial in $\xi_a$ and $\xi_r$. 

The results obtained here can be extended to a system of $N$ particles (ions), confined in a quadrupole electromagnetic trap with cylindrical symmetry \cite{Gheor2000, Major2005, Werth2009}. 
As shown in these references, the Hamiltonian of the center of mass (CM) is similar to the Hamiltonian for one particle $H$, given by Eq. (\ref{quadyn6}). Then the quasienergy 
states and the coherent states are exactly those for the $H$ Hamiltonian, where we have to substitute $M$ the mass of one particle with $NM$. In case when the interaction potential 
is translation invariant and homogeneous of degree $-2$, we obtain soluble models (e.g., for Calogero type potentials). These models enable to explicitly obtain bases and systems 
of symplectic coherent states, in order to investigate ordered systems made of trapped ions.

\section{Conclusions}\label{Concl}

The widest class of quantum models which describe both analitically and numerically the energy spectra and the solutions of the Schr\"odinger equation, is made of elementary 
quantum systems (in the Wigner sense). An elementary quantum system is described by an irreducible unitary representation ${\mathcal U}$ of a dynamical group ${\mathcal G}$ on a 
Hilbert space ${\mathcal H}$. The space of states is the quantum phase space ${\mathcal P}({\mathsf H})$, with the symplectic structure induced by the transition probability. It is 
assumed that the representation ${\mathcal U}$ admits coherent states. Then, coherent states manifolds are orbits of the action of the dynamical group on the space of states. The 
quantum Hamiltonian is a polynomial in the generators of the Lie algebra ${\mathfrak g}$ of the group ${\mathcal G}$. This property is specific to dynamical groups considered as 
spectrum generating groups. The energy (Husimi) function associated to the quantum Hamiltonian $H$ is a classical type Hamiltonian, whose values are precisely the expected values of the quantum Hamiltonian on 
coherent states. 

Therefore the quantum dynamical system is considered as a Hamiltonian system over the phase space of quantum states, while the corresponding classical system is described by the energy 
(Husimi) function over the phase space ${\mathcal M}$, whose points parametrize the coherent states. The symplectic structure of the classical phase space and the Hamilton 
equations of motion are explicitly obtained using the variational priciple applied on coherent states and from the theory of group representation.  
An algorithm has resulted, through which an energy 
function is associated to the quantum Hamiltonian. This function represents a classical type Hamiltonian, whose values are precisely the expected values of the quantum Hamiltonian. 
We suggest that this algorithm can be used to establish a correspondence between semiclassical and classical dynamics.

Semiclassical dynamics of trapped ions has been investigated by introducing the coherent state orbits as sub-manifolds of the quantum states' space with a K\"ahler structure induced 
by the transition probability. In case of dynamical symmetry groups, the solutions of the Schr\"odinger equation associated to linear Hamiltonians (in the infinitesimal generators 
of a dynamical symmetry group) are given by coherent vectors multiplied by geometric (Berry) phase factors for a confined ion or for the center of mass (CM) of a system of 
identical ions. Thus we obtain an energy spectrum for the time independent Hamiltonians (the case of Penning type traps) or a quasienergy spectrum for time periodical Hamiltonians 
(in case of dynamical Paul traps). The coherent states are parametrized by the solutions of the classical equations of motion. 

These results have been extended to the case of an ion confined in a quadrupole electromagnetic trap with anharmonicities (a situation according to reality), in order to study 
semiclassical dynamics associated to ions. Identification of equilibrium configurations is of interest for quantum information processing. We have obtained the explicit expression 
of the quantum Hamiltonian associated to a charged particle, located in an nonlinear electromagnetic trap (with anharmonicities), as a function of the Lie algebra generators for 
the radial and axial symplectic groups. The Schr\"odinger equation solution was found for the Hamiltonian we considered. An analytical potential with cylindrical symmetry was chosen. 
The Hamiltonian was particularized to the case of combined, quadrupole and octupole traps. In case of the pseudopotential approximation, the minimum points of the dequantified 
Hamiltonian define the equilibrium configurations for trapped ions, which enable implementing and scaling of quantum logic for larger number of ions.  

Although the Schr\"odinger equation is nonlinear, the Hamiltonian (Husimi) function generally determines nonlinear equations of motion which can be investigated using the theory of 
differential dynamical systems. This theory enables qualitative global investigation of Hamilton families, depending on various control parameters. In particular it allows to 
describe classical chaos using Morse theory, bifucation theory and catastrophe theory. Thus quantum chaos is described using the methods of study for the deterministic classical chaos. This description is possible 
through implicit codification in the Husimi function of the spectral properties of the Schr\"odinger operator. 
Dequantization by means of coherent states enables to achieve exact connexions between symmetrical quantum and classical dynamical systems. For an ion confined within a quadrupole 
trap, the requantification corresponds to a geometrical quantification.

\section{Acknowledgements}
The author would like to acknowledge support provided by the Ministery of Education, Research and Inovation from Romania (ANCS-National Agency for Scientific Research), contracts 
CEEX 05-D11-55/2005-2007 and PN09.39.03.01. The author (B. M.) also acknowledges fruitful discussions with Prof. Dr. Viorica Gheorghe, Dr. A. Gheorghe and Prof. Dr. M. Apostol.

\end{document}